\documentclass[conference]{IEEEtran}
\IEEEoverridecommandlockouts
\usepackage{amsmath,amssymb,amsfonts}
\usepackage{algorithm,algorithmicx,algpseudocode}
\usepackage{textcomp}
\usepackage{xcolor}
\usepackage[hang,flushmargin]{footmisc}
\usepackage{multirow}
\usepackage{booktabs}
\usepackage{graphicx}
\usepackage[colorlinks, urlcolor=blue]{hyperref}
\makeatletter
\renewcommand{\ALG@beginalgorithmic}{\small}
\makeatother
\newcommand\blfootnote[1]{%
  \begingroup
  \renewcommand\thefootnote{}\footnote{#1}%
  \addtocounter{footnote}{-1}%
  \endgroup
}

\begin{document}
% Pioplat: A Feasible and Customizable Latency Reduction Framework for Blockchain
% Pioplat: Win Latency War in Blockchain

% \title{Pioplat: A Feasible and Customizable Latency Reduction Framework for Blockchain}
%\title{Pioplat: A Distributed Framework for Winning the Blockchain Latency War}
\title{Pioplat: A Scalable, Low-Cost Framework for Latency Reduction in Ethereum Blockchain}
% 作者信息，为了匿名，需要隐藏掉。
%\author{Ke Wang}

\author{
\IEEEauthorblockN{Ke Wang, Qiao Wang, Yue Li, Zhi Guan, Zhong Chen}
\IEEEauthorblockA{\textit{School of Computer Science, Peking University,} Beijing, China}
\vspace{-20pt}
}

% \and
\maketitle
\pagestyle{plain}

\begin{abstract}
As decentralized applications on permissionless blockchains are prevalent, more and more latency-sensitive usage scenarios emerged, where the lower the latency of sending and receiving messages, the better the chance of earning revenue. 

To reduce latency, we present Pioplat, a feasible, customizable, and low-cost latency reduction framework consisting of multiple relay nodes on different continents and at least one instrumented variant of full node. The node selection strategy of Pioplat and the low-latency communication protocol offer an elastical way to reduce latency effectively. We demonstrate Pioplat's feasibility with an implementation running on five continents and show that Pioplat can significantly reduce the latency of receiving blocks/transactions and sending transactions, thus fulfilling the requirements of most latency-sensitive use cases. Furthermore, we provide the complete implementation of Pioplat to promote further research and allow people to apply the framework to more blockchain systems.
\end{abstract}


\blfootnote{
\kern -3pt
\hrule width 2in
\kern 2.6pt
The complete implementation of the Pioplat system is open-sourced and publicly available at \url{https://github.com/wangtsiao/pioplat}.
}

\section{Introduction}

Over recent years, permissionless blockchain supporting smart contracts, which basically anyone can access and develop applications on it, has attracted great attention. For a long time, the latency in blockchains has not been a first-order bottleneck of practical applications. Few works focused on latency reduction in existing blockchain protocols. However, the recent boom of on-chain decentralized financial (DeFi) applications brings attention to the latency war, where participants compete with each other for finance gains\cite{zhou2021high,wang2022cyclic}. In addition, regarding one of the intrinsic issues in permissionless blockchains: the maximal extractable value (MEV), where miners, bots, and traders all try to make more profits from block production by including, excluding, and changing the order of transactions in a block. Out of these, participants with lower latency have a more significant opportunity of earning benefits. Thus all the above make there a latency war happening silently in the underlying p2p network of blockchains\cite{tang2022strategic,babel2022strategic}.

Traders invest heavily in laying fiber or satellite links in traditional financial markets to reduce latency by a few milliseconds\cite{osipovich2020high,osipovich2021high}. However, in a decentralized blockchain, where no centralized institution and nodes broadcasting transactions/blocks are spread worldwide, the traditional method of reducing latency is not feasible. There are some paid services, such as bloxRoute\cite{bloxroute_2022}, whose paying users utilize dedicated gateways to stay connected with those services' private nodes, which are connected via the cut-through private network. However, their mechanism for reducing latency is not publicly available. The critical shortcoming is that users cannot customize their latency reduction preferences elastically. The distribution of these private nodes is not known, and these paid services have the ability to censor transactions/blocks. In addition, most of these services have usage restrictions for commercial reasons, such as limiting the number of transactions sent, duration of connectivity, etc. In this work, we are dedicated to providing a feasible, customizable, and publicly accessible framework, thus helping users win the blockchain latency war.

Blockchain is built on the unstructured p2p network where nodes are randomly connected to each other\cite{wang2021ethna}. The intuitive idea is to modify the p2p topology for faster block and transaction propagation. e.g., a structured p2p overlay as a faster alternative to the random topology\cite{rohrer2019kadcast}. However, this idea does not apply to existing blockchains because we can not modify the blockchain protocol. Since the goal is to compete with others for lower latency, a independent solution of the blockchain protocol is needed. Perigee found that connecting different neighbors in a p2p network can have a hundred-millisecond magnitude of latency impact\cite{mao2020perigee}. Perigee designed a neighbor selection strategy that adaptively selects nodes in the p2p network purely based on interactions with neighbors. Perigee scores its neighbors by examining the message arrival times, where neighbors who consistently deliver messages quickly are favored while others are disconnected. But Perigee has an issue that most of the neighbors after adaptive selection are nodes that are geographically closer, and in practice, we have found this fact. Since blockchain is a worldwide system, selecting geographically close neighbors is not what we expected, and this issue of Perigee will cause us to miss messages on other continents. Therefore, it can not fulfill the requirement of blockchain latency reduction. Choosing Perigee means giving up on connecting to nodes on other continents. A na\"ive approach uses multiple nodes equipped with a neighbor selection strategy on different continents, and these nodes prioritize the delivery of messages to each other. However, this ignores that running nodes that can receive complete block/transaction objects require a lot of storage and powerful computing performance. This na\"ive solution is not practical but it provides a good idea.

We present Pioplat, a feasible, low-cost, customizable latency reduction framework that allows users to elastically win the blockchain latency war, consisting of an instrumented variant of full node and multiple relay nodes on different continents. We avoid modifying blockchain protocols on which the community has already reached a consensus; Instead, we adaptively select neighbors and deploy multiple relay nodes in a distributed manner to reduce latency in a non-intrusive way. The user of Pioplat can elastically add or remove relay nodes to make a trade-off between their latency requirements and expenses. Because miners generally congregate in certain geographic areas\cite{silva2020impact}, we provide a tunable parameter allowing the user to customize neighbor selection preference. This can make a relay node located in a miner-dense area prefer neighbors that deliver blocks fast. In contrast, relay nodes in other regions favor neighbors that relay transactions fast. Our contributions are as follows:
\begin{itemize}
  \item To the best of our knowledge, Pioplat is the first publicly available, blockchain latency reduction distributed framework. We show that it can reduce latency by elastically adding low-cost relay nodes.
  \item We present a neighbor selection strategy that allows relay nodes to simultaneously select neighbors delivering blocks quickly and neighbors delivering transactions quickly.
  \item We carefully design a low-latency communication protocol between relay nodes, suitable for messages in blockchain. We also describe some optimizations applied in our implementation.
  \item We conducted an extensive evaluation to demonstrate Pioplat's performance in blockchain latency reduction. We give open-source tools that allow other researchers to replicate our work and conduct similar research in more blockchain systems.
\end{itemize}

\section{Background}
For the sake of concreteness, we focus on the Binance smart chain in this work, which is a leading Ethereum-similar blockchain system\cite{cernera2022token}. It should be noted that this does not mean that Pioplat proposed in this paper is not applicable to other blockchains; On the contrary, Ethereum and Ethereum similar blockchain systems such as Torn and Polygon can all benefit from the Pioplat framework. Users of other blockchains can also reduce latency  according to the ideas presented in this work. In this section, we give details on the information propagation mechanism in blockchain (\ref{background_subsection_A}), the respective reason behind the latency of a client receiving transactions/blocks and sending transactions (\ref{background_subsection_B}), and several latency-sensitive use cases at the time of writing (\ref{background_subsection_C}).

% \cite{mokalusi2023factors}

\subsection{How messages are propagated}
\label{background_subsection_A}
Blockchain uses a gossip-similar strategy to propagate blocks/transactions in the underlying p2p network\cite{wang2021ethna}. No matter what consensus algorithm is adopted, miners broadcast the block newly mined directly to all neighbors to notify nodes in the network as soon as possible to enter the next block production cycle. To reduce the traffic redundancy of the gossip strategy, nodes in the blockchain relay the complete object of blocks/transactions or just the announcement that contains only the hash of the object. Once a node receives a block never seen before, it first checks validity, such as signature, root hash, etc. Then the node selects neighbors that do not have the block, which is determined by whether the neighbor has sent the block or the announcement of the block to the node. Further, the node randomly picks a small set from such neighbors to directly send the complete block object. The rest of the neighbors are sent the block's hash as an announcement indicating the node already has the corresponding block. Once a node receives an announcement of a block never seen before, it first waits a while, 300-500 ms subject to users, if it still did not receive the complete block object, then randomly picks a node $A$ from among neighbors who have already announced. After that, the node requests the block header to $A$, and after receiving the header, the node checks its validity. If the header is correct, then the node requests the block body to $A$ otherwise drops neighbor $A$ and repeats the header request to other neighbors. Once the node gets the complete block object, it starts propagating the block and the announcement to neighbors who still do not have the block. 

Regarding transaction propagation, when two nodes have established a connection, they should exchange each other's transactions in their local transaction pool, which is subject to client-specific limits and contains many pending transactions. This way helps deliver those pending transactions to miners who further pick them for inclusion into the blockchain. Specifically, both nodes send announcements to each other containing the hashes of all transactions in the transaction pool. The node filters the received set and picks transaction hashes that it does not have in its local transaction pool. After then, it requests another node for the complete transaction object. Once a node receives the object, it makes the same actions as block propagation. It propagates the complete transaction object to a small random set of neighbors that do not have the transaction and propagates the corresponding announcement of the transaction to the rest neighbors. Thanks to the mechanism above, there are no nodes that relay the same block or the same announcement to a neighbor which previously has the block, thus reducing the workload of the network. However, the mechanism obviously introduces a significant latency, which we will further discuss (\ref{background_subsection_B}).

\subsection{What led to the latency}
\label{background_subsection_B}
In this work, we do not consider the inherent latency of the network transport layer. This is because what actually has a greater impact on blockchain latency is p2p topology, the propagation mechanism above, and the time-consuming validity verification\cite{mao2020perigee}. 

For p2p topology, blockchain systems utilize a random connection policy to achieve better decentralization to form the underlying p2p network. In particular, in such a worldwide public network, this random connection is likely to cause two nodes $u$ and $v$ which are geographically close to being connected by an intermediate node $w$ that is distant from them. This obviously greatly prolongs the message propagation delay between $u$ and $v$. Prior works have extensively investigated p2p neighbor selection strategies; it has been proved that finding the global optimal topology of blockchain p2p network using neighbor selection is an NP problem \cite{tang2022strategic}. However, it does not mean that the issue above is completely unsolvable; the prior work has pointed out that a simple neighbor selection strategy, which will be further discussed and instrumented (\ref{design_subsection_B}), although can not find the global optimum, can reduce latency well\cite{tang2022strategic}. 

For time-consuming validity verification, after each node receives a block or transaction, the node verifies its correctness before sending the complete object or announcement to its neighbors. Although a single validation duration is short compared to round-trip-time between nodes, a block or transaction may go through several times validation from being sent to being received by a node, this inevitably makes the overall latency that transmitted blocks/transactions experience worse. 

For propagation strategy, the first concern as a node is to receive the complete objects of block/transaction directly from neighbors as soon as possible, rather than receiving an announcement and then requesting the corresponding object as we described above. Obviously, the latter greatly prolongs the latency. Further, regarding the latency of sending transactions, if a node sends a transaction, this transaction is added to the local transaction pool, and then the $\sqrt{N}$ of the N neighbors are randomly selected to deliver the completed transaction object to them, delivering the announcement to the rest. Such a sending transaction strategy makes the transaction take a longer duration to reach miners. Pioplat reduces the duration by increasing the number of neighbors and broadcasting the complete transaction object directly to all of them. We will further detail strategies adopted by Pioplat to reduce latency.

\subsection{Latency-sensitive use cases}
\label{background_subsection_C}
We present a few examples of latency-sensitive illustrations to emphasize the importance of latency reduction in blockchain. These examples all have a common feature, the lower the latency of the participants the more they profit.

\textbf{High frequency trading}. The most common form of high-frequency trading on blockchains is arbitrage, which can provide liquidity to traders. The concept is simple: find the spread between exchanges for a given token, send the arbitrage transaction, then profit by wiping out the token spread between exchanges in the process. Because a spreading opportunity allows only one arbitrage trade, in other words, it is mutually exclusive. Only the first trading transaction that reaches the miner can make a profit. Regarding the arbitrage computation problem, take arbitrage in AMM exchanges as an example, it was solved using the convex optimization algorithm since this part of the computation time has not been a bottleneck for competing for latency compared to the network transmission time\cite{noxx_2022}. The network transmission time is usually tens or even hundreds of times longer than the computation time.

\textbf{Building blocks}. To ensure that the vast majority of participants in a p2p network execute the transactions within a block in a single block production duration, the blockchain uses gas to measure the amount of computation required by a program to perform an action or set of actions and to limit the total amount of gas for all transactions in a single block. Gas price is the amount of money a user is willing to pay for a single unit of gas in his/her transaction. Miners, as producers of blocks, can gain more revenue if they include as many of those with higher gas price as possible within the limited total gas constraint\cite{liu2022empirical}. Reducing the latency in collecting transactions gives miners more time to make selections and, empirically, gives miners the opportunity to collect more transactions per unit of time and therefore have more options to make a profit\cite{mazorra2022price}.

\section{Design}
In this section, we give a complete overview of the whole system (\ref{design_subsection_A}), then present the customizable p2p neighbor selection strategy (\ref{design_subsection_B}), and next show how a client becomes indistinguishable through redirecting requests (\ref{design_subsection_C}) and finally discuss low-latency communication protocol between relay nodes (\ref{design_subsection_D}). We discuss how the following requirements are satisfied:
\begin{enumerate}
    \item[R1:] Reduce the latency of receiving blocks/transactions and in the meantime make sent transactions be confirmed early.
    \item[R2:] Reduce message delivery latency between relay nodes, and try to avoid network congestion.
    \item[R3:] Minimize the hardware requirements while making relay nodes work well.
\end{enumerate}

\begin{figure}[htbp]
\centerline{\includegraphics[width=0.45\textwidth]{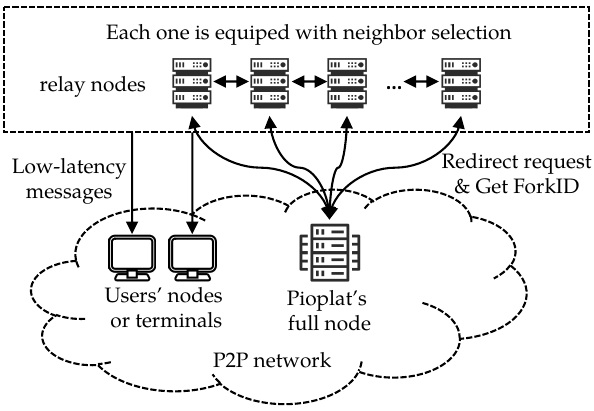}}
\caption{Overview of Pioplat system: relay nodes distributed across different continents are connected to each other via a low-latency communication protocol, and they are all connected to one (or more) instrumented variant full node to obtain the forkID and redirect data requests. Users of Pioplat reduce latency by simply connecting their full nodes to one or more relay nodes.}
\label{fig_pioplat_nutshell}
\end{figure}

\subsection{Pioplat in a nutshell}
\label{design_subsection_A}
In the p2p network, from the topological point of view, more neighbors bring the node closer to the miners and nodes who broadcast transactions frequently, and blockchain clients usually have a parameter $M$ that specifies the maximum number of neighbors that stay connected\cite{kim2018measuring}. Moreover, more neighbors also means a greater possibility of receiving complete objects in a limited time, sending transactions to more nodes, which allows these transactions to reach miners early. As we want to reduce the latency of receiving blocks/transactions and sending transactions, a natural idea is to increase the parameter $M$. However, this gain is not without downsides; more neighbors take up larger bandwidth and increase the likelihood of traffic congestion which may cause a lot of TCP packet retransmissions, making latency even worse. The single-node approach is not scalable and cannot fulfill geographically distributed requirements of the blockchain. Pioplat is composed of distributed multiple nodes spread across different continents. Specifically, Pioplat has two kinds of nodes: the instrumented variant of full node client which persists blockchain data and syncs to the latest block height, and the relay node which only implements blocks/transactions propagation protocol and does not store blockchain data. When a relay node establishes a connection with a new neighbor or is requested to provide data by a neighbor, the full node is responsible for providing information to the relay node, e.g., the genesis hash, the forked block number, the current block height, etc. Each relay node performs a neighbor selection strategy in the blockchain p2p network purely based on timestamps of receiving blocks/transactions. These relay nodes form a scalable, low-latency transaction and block dissemination network. Currently, most blockchain clients support adding trust nodes, which means that even if a corrupted message is received from any node in the trust node list, the client continues to stay connected with them\cite{eth2014geth,bnb2022bsc,ledger2020erigon}. So the relay node does not check the validity of blocks/transactions, it simply acts as a gateway to directly deliver messages to other nodes that have validation ability. To reduce latency, users just add one or several relay nodes to the trust node list and connect their blockchain clients to those nodes.

\textbf{Pioplat utilizes the cache to be an affordable solution.} Pioplat eliminates the expensive storage cost, and even an introductory instance provided by a cloud service vendor can run the relay node, making the system affordable for users.  Our design, shown in Fig.\ref{fig_pioplat_nutshell}, sets up a cluster consisting of several relay nodes. Usually, a blockchain full node occupies a lot of storage space. At the time of writing, running a full node of the \textit{bsc mainnet} requires a fast SSD drive with at least 1.7TB of space, and the storage requirement will continue to grow over time. Thus, it is impractical to directly deploy multiple full nodes, which will cause significant expense and waste of resources. Pioplat provides a solution where theoretically only one full node is required, and more can be added in practice for fault tolerance. The role of the full node will be described in the following. The magic that reduces huge storage requirements is the block cache, the transaction cache, and request redirecting. These two caches are used to replace the mass storage in the normal full node to serve most of the requests coming from neighbors. 

The block cache keeps a fixed-length block in the order of block number regardless of whether multiple blocks are generated at the same height, which ensures that forks occurring in the network are monitored quickly. As shown in Alg.\ref{algo_fixed_size_block_cache}, $K$ represents the maximum number of blocks containing unique numbers that in the cache, $i$ stands for the index to access block number list $\Phi$ that contains numbers of the currently cached blocks, $\Lambda$ acts for a key-value hash map that it allows to access blocks with the number in $O(1)$ time. After receiving a block, we first check whether there have other blocks at the same height if it does then we call the block a fork block (ln.2,3) and check whether the block is already in the cache (ln.4-9). If it is a fork block then we append it to $\Lambda$ and directly return (ln.11-14). Otherwise, we save the block (ln.19,20), clear history blocks (ln.15-17), and update the index (ln.21).

% 逐行的介绍算法（或者逐块的介绍）

\begin{algorithm}
        \renewcommand{\algorithmicrequire}{\textbf{Input: }}
        \algorithmicrequire A block $B_{n,h}$ with number $n$, hash $h$.
	
        \caption{Enqueue Block to the Cache}
	\label{algo_fixed_size_block_cache}
        \begin{algorithmic}[1]
            \State Init $shouldInsert, isForkBlock, i \leftarrow true, false, 0$
            \If{$n$ in $\Lambda$} \Comment{Check whether $B_{n,h}$ is a fork block}
                \State Set $isForkBlock \leftarrow true$
                \For{each block in $\Lambda[n]$}
                    \If{$h$ equal hash of block}
                        \State Set $shouldInsert \leftarrow false$ \Comment{Already enqueued}
                    \EndIf
                \EndFor
            \EndIf

            \If{$shouldInsert$ equal $true$}
                \If{$isForkBlock$ equal $true$}
                    \State Append $B_{n,h}$ to $\Lambda[n]$ \Comment{Store fork blocks}
                    \State \textbf{return}
                \EndIf

                \If{Count of $\Lambda$ keys $\geq K$}     \Comment{Remove previous blocks}
                    \State Block number to remove $r \leftarrow \Phi[(i+K-1) \ mod \ K]$
                    \State Delete all blocks of $\Lambda[r]$
                \EndIf
                \State Set $\Phi[i] \leftarrow n$       \Comment{Write the block number}
                \State Append $B_{n,h}$ to $\Lambda[n]$ \Comment{Save the block to cache}
                \State Set $i \leftarrow (i + 1) \ mod \ K$  \Comment{Update the index}
               \State \textbf{return}
            \EndIf
        \end{algorithmic}
\end{algorithm}

Considering the fluctuation of user activity and the diurnal transaction traffic patterns, one should not simply set a fixed size for the transaction cache, otherwise, transactions exceeding the cache limit will be discarded when there is a large number of transactions in the network. Pioplat uses a time window to maintain the transaction cache. As shown in Alg.\ref{algo_transaction_cache_with_expire_duration}, $\Upsilon$ represents a key-value hash map that allows access to the transaction with a hash in $O(1)$ time. $\Psi$ denotes a priority queue, each element of which is a transaction hash and a timestamp when this transaction is added. When removing elements, $\Psi$ prioritizes removing the element with the smallest timestamp. Record the timestamp of each transaction when it is added to the cache (ln.2,3), and delete those that have expired in the cache for more than $Expire$ whenever a new transaction is inserted (ln.5,6).

\begin{algorithm}
	\renewcommand{\algorithmicrequire}{\textbf{Input:}}
        \algorithmicrequire A transaction struct $T_h$ with hash $h$.

        \caption{Put Transaction to the Cache}
	\label{algo_transaction_cache_with_expire_duration}
	\begin{algorithmic}[1]
		  \If{$h$ not in $\Upsilon$} \Comment{Check whether $T_h$ exists}
                \State Set $\Upsilon[h] \leftarrow T_h$
                \State Add $(h, Expire)$ to $\Psi$ \Comment{Record expiration time}
            \EndIf
            \For{$h$ in $\Psi$'s expired transactions hash list}
                \State Delete $\Upsilon[h]$
            \EndFor
	\end{algorithmic}  
\end{algorithm}

\textbf{Pioplat relies on neighbor selection.} Pioplat adopts an approach that sets different neighbor selection preferences for each relay node based on the nodes' distribution at the region. Today, the underlying p2p network of blockchain systems deploys the simple random connection policy\cite{kim2018measuring}, where there may be two nodes, while the shortest path between them passes through another intermediate node that is geographically distant from the two nodes. This makes message propagation latency awful. To further complicate matters, now we are facing the polycentricity of miners such as mining pools and extensive use of cloud computing\cite{wang2021ethna}. In addition, even though p2p networks are censorship-resistant, network censorship of cryptocurrencies in some regions has significantly prolonged their's latency from the outside world\cite{han2022using}. This means that the distribution of miners and nodes which frequently broadcast transactions varies in different regions. The relay node recognizes the speed of message propagation from its neighbors by the timestamp of the received message. Suppose many miners but few nodes that broadcast transactions are gathered in a wide area, this is possible because the miners favor lower electricity prices. In this case, the neighbor selection preference of the relay node tends to select nodes that deliver blocks earlier. Vice versa the selection preference tends to choose nodes that deliver transactions earlier. Pioplat introduces a parameter that adjusts the degree of influence of these two preferences. Thus satisfying the R1.

\textbf{Pioplat propagates messages between relay nodes.} As a system dedicated to reducing blockchain message latency in a geographically distributed manner, low-latency communication between relay nodes is critical. Blockchain systems today rely on the RLPx protocol, a secure transport protocol on top of TCP protocol to transmit encrypted blocks and transactions in the underlying p2p network\cite{eth2014geth,bnb2022bsc,matic2020bor}. Each p2p node spends an amount of time cryptographically verifying the authenticity of receiving blocks/transactions from neighbors, which varies depending on the computational power of the node. We design a low-latency communication protocol between relay nodes with a combination of TCP and UDP. According to the aforementioned trust nodes mechanism of blockchain client, Pioplat eliminates validation of the authenticity of blocks/transactions thus skipping the costly computation processes. Finally, we employ several optimizations to make a trade-off between UDP reliability and bandwidth utilization, and to avoid duplicate en-/decoding and en-/decryption of messages, further saving time.

\subsection{Neighbor selection}
\label{design_subsection_B}
Perigee is an adaptive approach to select neighbors, it is purely based on a node's historical interactions with its neighbors to determine which neighbors it should stay connected\cite{mao2020perigee}. Inspired by the Perigee, Peri added a permanent blocklist mechanism on top of Perigee\cite{tang2022strategic}. We have to admit that Perigee and Peri are outstanding neighbor selection solutions. However, we found two shortcomings in them. First, Perigee was not originally designed to reduce latency but to increase the throughput of the blockchain system, it only considers reducing the latency of receiving blocks. Peri was designed to only reduce the latency of receiving transactions. They ignore that reducing the latency of receiving transactions and receiving blocks are both essential to latency-sensitive use cases. Pioplat improves these algorithms by introducing an adjustable parameter that simultaneously reduces blocks and transactions receiving latency. The details will be described below. Second, they both did not consider the churn issue in the p2p network. A neighbor was removed may be due to its own high latency, temporary unexpected network conditions, or user interruptions. To preserve valuable slots of p2p neighbors' connection, the high latency nodes should no longer be allowed to reconnect, but for not missing any low-latency neighbors, the other neighbor dropped due to temporary reasons should be allowed to reconnect. Pioplat mitigates this problem well with an expirable blocklist mechanism which will be explained below.

Based on Perigee\cite{mao2020perigee}, we designed the following neighbor selection strategy. The relay node regularly checks how quickly blocks and transactions are delivered from each of its neighbors and then decides whether to keep them as neighbors or drop them and look into connecting to potentially more effective nodes. Alg.\ref{algo_neighbor_selection} is the pseudocode of the neighbor selection algorithm. We use $K$ to denote the maximum number of neighbors allowed to stay connected, which is determined by the user according to the performance and bandwidth of the node; $\rho$ denotes the ratio of neighbors removed in each round; $\sigma$ denotes the ratio of neighbors to reserve who deliver blocks quickly; then $(1 - \rho - \sigma)$ is the ratio of neighbors to reserve who deliver transactions quickly. In regions with fewer miners but many nodes broadcasting transactions, such as East Asia, set $\sigma$ smaller, and vice versa set it larger. $\Psi_{tx}$ is a key-value map dedicated to recording transaction information, where the key is the transaction hash and the value is a list of items, each of which is a neighbor identifier $p$ and a timestamp when the message was received. The structure of $\Psi_{block}$ is the same as $\Psi_{tx}$, the difference is that it records timestamps of receiving blocks. Whenever a transaction or block message is received from a neighbor $p$, either a hash announcement or a message containing a complete object, the timestamp is recorded using \textit{onReceiveTransaction} (ln.14-16) and \textit{onReceiveBlock} (ln.17-19), respectively. Every $\Delta$ time, Pioplat computes two scores for each neighbor based on the timestamps of transactions and blocks arrival in the past time respectively. These scores are stored in two lists $S_{tx}, S_{block}$ respectively (ln.23,24). The \textit{getScores} function used here is directly derived from the Perigee strategy calculation method, and this paper will not dwell on it\cite{mao2020perigee}. After that, sort $S_{tx}, S_{block}$ and save the smallest $K*\sigma$ and $K*\epsilon$ respectively to get the union set of neighbors to be kept $P_{new}$ (ln.26-33). Then IPs of $P-P_{new}$ are added to the blocklist, and the expiration time is set to $2^n * 20$ minutes if $p$ has appeared in the blocklist $n$ times before (ln.34-41). The blocklist is set in this way mainly to give chances to reconnect nodes that have been dropped due to unexpected network conditions or user interruption.

\begin{algorithm}
        \algdef{SE}[DOWHILE]{Do}{doWhile}{\algorithmicdo}[1]{\algorithmicwhile\ #1}%
        \caption{Neighbor Selection in p2p Network}
	\label{algo_neighbor_selection}
	\begin{algorithmic}[1]
            \Function{init}{}
            \State set $B \leftarrow \varnothing$ \Comment{Expired blocklist}
            \State set $P \leftarrow \varnothing$ \Comment{Neighbors}
            \State set $\Psi_{tx},\Psi_{block} \leftarrow \varnothing,\varnothing$
            \EndFunction
\item[]
            \Function{onConnectNewPeer}{$p$}
                \If{$B$ contains $p$ and $B[p]$ is not expired}
                \State refuse this connection
                \Else
                \State accept this connection
                \State $P \leftarrow P + p$ \Comment{Add $p$ to neighbors}
                \EndIf
            \EndFunction
\item[]
            \Function{onReceiveTransaction}{$hash$, $peer$}
                \State append $(peer, timeNow)\  to\ \Psi_{tx}[hash]$
            \EndFunction
\item[]
            \Function{onReceiveBlock}{$hash$, $peer$}
                \State append $(peer, timeNow)\  to\ \Psi_{block}[hash]$
            \EndFunction
\item[]
            \Function{Start}{$K$, $\rho$, $\sigma$, $\Delta$}
                \State set $\epsilon \leftarrow 1-\rho-\sigma$
                \Do
                    \State sleep $\Delta$ seconds
                    \State set $S_{tx}, S_{block} \leftarrow getScores(\Psi_{tx}),getScores(\Psi_{block})$
                    \State set $\Psi_{tx},\Psi_{block} \leftarrow \varnothing,\varnothing$
                    \State set $P_{new} \leftarrow \varnothing$ \Comment{Neighbors to be retained}
                    \State set $P_{new} \leftarrow P_{new} \cup smallestN(S_{tx}, K*\sigma)$
                    \State set $P_{new} \leftarrow P_{new} \cup smallestN(S_{block}, K*\epsilon)$
                    \If{$|P_{new}|>K*(1-\rho)$}
                        \State set $t \leftarrow |P_{new}| - K*(1-\rho)$
                        \State set $r \leftarrow smallestN(S_{tx}, t) \cup smallestN(S_{block}, t)$
                        \State set $P_{new} \leftarrow P_{new} - randomK(r, t)$
                    \EndIf
                    \For{$p$ in $P-P_{new}$}
                        \State set $t \leftarrow timeNow()$
                        \If{$p$ has appeared in $B$ for $n$ times} 
                            \State set $B[p] \leftarrow t + (20*2^n) \ minute$
                        \Else
                            \State set $B[p] \leftarrow t + 20 \ minute$
                        \EndIf
                    \EndFor
                \doWhile{$continue == true$}
            \EndFunction
	\end{algorithmic}  
\end{algorithm}

\subsection{Communication between relay nodes}
\label{design_subsection_D}
Relay nodes receive messages from the network with lower latency since they are equipped with the neighbor selection that favors nodes delivering blocks and transactions early. However, limited by the computing hardware and bandwidth, a single relay node can only connect a restricted number of nodes. By default the Binance smart chain reference client allows a node to connect with at most 50 neighbors, which is a drop in the bucket compared to thousands of nodes in the network. Moreover, under the above neighbor selection strategy, each relay node tends to connect neighbors that are geographically near, which constrains it from receiving transactions or blocks on other continents with low latency. Therefore, we designed Pioplat as a distributed system spanning different continents. It contains multiple relay nodes, and users anywhere in the world can access any relay node to reduce the latency of receiving blocks/transactions and to make sent transactions confirmed faster. For relaying messages quickly, the communication protocol between the relay nodes is critical to latency reduction. In this section, we describe the detail of the communication protocol and how Pioplat minimized the overhead to achieve an elastic latency reduction system.

\begin{figure}[htbp]
\centerline{\includegraphics[width=0.5\textwidth]{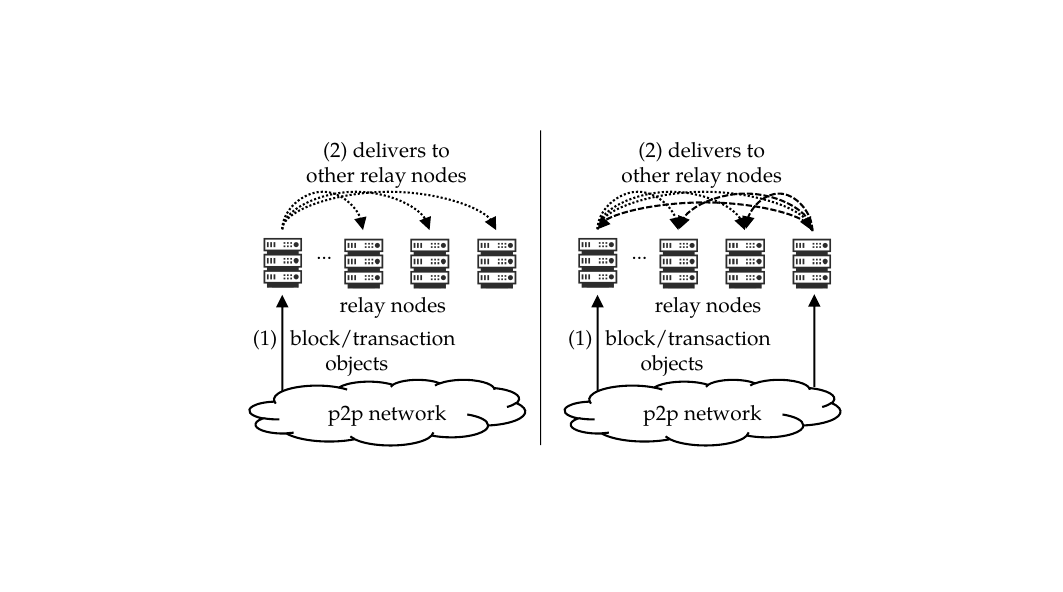}}
\caption{When a Pioplat relay node receives a block or transaction body message never seen before from the blockchain p2p network, the relay node broadcasts the message to other relay nodes at first. The left figure depicts the expected scenario; The right figure depicts the redundant message transmission between relay nodes when multiple receive new messages simultaneously.} // TODO
\label{fig_cluster_deliver}
\end{figure}

\textbf{Communication between relay nodes.} In the blockchain, there are two modes of blocks/transactions relaying. The first is relaying hash announcement, nodes receiving the announcement send requests to fetch the corresponding complete object. This mode introduces an extra round-time-trip delay. The second is relaying complete objects directly, since there is no need for the following request, the latency is lower. Unfortunately, a node by default has a high probability to select the first mode for relaying to its neighbors\cite{wang2021ethna}. On the one hand, the Pioplat relay node increases the possibility that a neighbor broadcasts complete objects to it in a limited time via the second mode by staying connected to more neighbors simultaneously. On the other hand, when a relay node receives a complete object from a neighbor and the object does not exist in the local cache of the relay node, the relay node then broadcasts this complete object directly to the other relay node, if the complete object is received from another relay node, the relay node just delivers it to users and its neighbors then put the corresponding block/transaction into the cache to serve requests from neighbors. If the block/transaction is already in the local cache then the relay node does nothing. It is worth mentioning that if two or more relay nodes simultaneously receive a block or transaction that has never been seen, those nodes will concurrently broadcast the block or transaction to all other relay nodes except itself. Although this causes duplication of transmission, we argue that this is unavoidable and necessary to achieve low-latency delivery of blocks/transactions. Since the introduction of any coordinator will undoubtedly increase the latency, we have no choice but to let the relay node that first monitors the message immediately relay.

\textbf{Using UDP to deliver small transactions.} The time interval for generating blocks tends to be roughly constant in a well-working blockchain system. It is about 3 seconds for Binance Smart Chain, 12 seconds for Ethereum, and 2.3 seconds for Polygon\cite{bnb2022bsc,eth2014geth,matic2020bor}. In other words, the number of blocks propagating in a blockchain p2p network does not change much over some time and the size of a block is usually around 20kB. Since the traffic is stable, the TCP, as a reliable, connection-oriented protocol is suitable for the transmission of blocks. However, this is not the case with transactions. User activity is not evenly distributed over time. e.g., there are fewer transactions when the market is bearish and a much higher number of transactions when the market is bullish. If transactions are transmitted using the TCP, network congestion will trigger retransmission, further causing a jitter that can dramatically increase latency. We statistically measure the size of transactions encoded in RLP, which is an encoding method used when they are transferred in the network, on three leading blockchains' mainnet for 24 hours, and the result shows that the 99.4\% of transactions RLP-encoded size is no more than 1432 bytes\footnote{Maximum transmission unit subtract the size of the IP/UDP header and size of the header in the datagram, which will be described further in section.\ref{section_impl}.} in Binance smart chain, 97.4\% in Ethereum, and 97\% in Polygon. Compared to TCP protocol UDP has no concept of acknowledgment, retransmission, and timeouts. All these features make UDP a low-latency transport layer protocol. It is suitable for latency-sensitive use cases\cite{wiki:User_Datagram_Protocol}. Although UDP is an unreliable protocol, the fact that a few transactions do not reach does not affect performance. Moreover, the redundant broadcast of Pioplat when multiple relay nodes receive the same transaction simultaneously also mitigated the unreliability problem of UDP. The size of a transaction is much smaller than that of a block, and relay nodes deliver transactions that can be accommodated in a network package to each other using UDP. This satisfies the R2.

\subsection{Handshaking and redirecting requests}
\label{design_subsection_C}
In Pioplat, we try to eliminate full nodes by using relay nodes, which, as described in \ref{design_subsection_A}, cache blocks and transactions temporarily in the local and do not store blocks and state persistently, just requiring only a small amount of hard disk storage to persist log and elite neighbors. However, since the relay node does not persist data, it is a challenge to let normal full nodes in the blockchain p2p network treat the relay node as a full node in their view. Because only then can the nodes in the p2p network stay connected to the relay node\cite{kim2018measuring}. In addition, ethically we do not expect relay nodes to have a negative impact on the decentralized blockchain network, so making these relay nodes have the ability to serve protocol-compliant requests from neighbors should be considered. In this section, we describe how Pioplat addresses these two challenges and how Pioplat becomes a feasible latency-reduction framework that is friendly to the decentralized blockchain protocol meantime.

\textbf{Get information for handshaking with neighbors.} Today's blockchain p2p networks are filled with a wide variety of clients which run different blockchains with the same implementation. Including public or private, such as Ethereum mainnet and Ethereum private network; different blockchain systems, such as Binance smart chain and Ethereum; different hard-forked networks of the same blockchain, such as Ethereum and Ethereum classic. It is important to identify the neighbors that should be connected in the intricate network, the p2p handshake mechanism is here to handle this issue. The current implementation uses genesis hash, the current latest block height, and the list of forked block numbers with corresponding hashes to identify the network\cite{mohammed2021hyperledger}. Of course, sending all the information above to the other side would be inefficient during the handshake. The fork-id mechanism proposed by EIP-2124 compresses the above information into a 4-byte message, efficiently handling the handshake. Relay nodes request the current fork id to Pioplat's instrumented variant of a full node to establish a connection with the neighbor. Among the source data for generating fork ids, the genesis hash keeps unchanged for all time, hard forked happens accompanied by clients upgrade which is not usual, and only the current latest block height is changed every several seconds. This means that the fork id is unchanged during a block-producing interval. It is not necessary that a relay node requests fork id from the full node each time connecting to a new neighbor. In order to reduce the communication overhead, the fork id is cached by the relay node, and each time it is used, the relay node first checks if it is more than one block-producing interval since the last cache, if it is, then the relay node requests the full node for a new fork id and update local cache, otherwise, relay node directly uses the fork id in the cache. This approach is effective and reduces communication costs.

\textbf{Redirect requests from neighbors to the full node.} The full node is expected to have knowledge of the complete chain of all blocks from the genesis block to the current latest block. By the mechanism described above, a node $N$ in the blockchain network cannot distinguish our relay node from the regular full node except the node $N$ request data that does not exist in the local cache of the relay node. This exception is typical and can hurt latency if left unaddressed\cite{kim2018measuring}. For example, suppose a neighbor tries requesting a block announced by our relay node but does not receive a response containing the corresponding block before timeout or receives an invalid response. In that case, the neighbor will actively disconnect from our relay node. Fewer neighbors will reduce the possibility of the relay node receiving complete block/transaction objects directly in a limited time. Pioplat handles this through redirecting requests. When a relay node receives a fetching data request, it responds directly if it has the corresponding data in the local cache. Otherwise, it redirects the request to the full node of Pioplat. After it responds, the relay node responds to the requester. In this way, the relay node acts as a proxy of the full node. Since the relay node satisfies all its neighbors' requests, the system does not negatively affect the decentralized blockchain network. Thus fulfilling the requirement R3.

% We do not choose Ethereum because it has recently completed the merge between PoW and PoS, which consists of a beacon chain and a main chain now, if we choose it that the subsequent experiments in this work require both modifications and deploying a consensus client and an execution client, which is a heavier workload and expensive server cost for us\cite{kapengut2022event}. Roughly speaking, except for the consensus mechanism, the Binance smart chain (bsc) is identical to Ethereum before the merge\cite{cernera2022token}. Transaction financial expenses of the Binance smart chain are lower, which is more suitable for experiments of this work.

\section{Implementation}
\label{section_impl}
This section describes the implementation of the relay node. Our goals are to let the relay node handshake well when it connects to new nodes and avoid neighbors disconnecting with the relay node actively. This fundamentally requires us to follow the specifications of the blockchain protocol strictly. Although the relay node does not process blocks and transactions, it relies on p2p communication and cryptography, so developing it from scratch still requires considerable effort. Naturally, using a prevalent client of these blockchains as a basis of implementation is a good choice. Based on the Go-Ethereum that the most pervasive client of Ethereum\cite{eth2014geth}, bsc is the reference implementation of Binance smart chain which is one of the leading permissionless blockchains supporting smart contracts\cite{bnb2022bsc,cernera2022token}. Another rationale behind taking bsc instead of the other client is that Golang is a language suitable for network programming because it has a better peculiarity of concurrency support. In this section, we present system components and several optimizations employed in Pioplat. 

\subsection{System components}
\textbf{Sending transactions:} In the current blockchain client, such as bsc, if a user sends a transaction, the user first submits the transaction to a blockchain client through a remote procedure call. Then the blockchain client adds the transaction into its transaction pool and checks signature correctness, nonce continuity, etc. If these checks pass, then the transaction is placed in the transaction queue. The blockchain client periodically takes transactions from the transaction queue and broadcasts them to its neighbors. If there are $n$ neighbors to broadcast, the client randomly picks $\sqrt{n}$ neighbors from those $n$ neighbors to send the complete transaction object and the rest to send the corresponding hash announcement. Those who received only hash announcement need to request further the complete object, which prolongs the duration between the user sending the transaction and the miners who produce the next several blocks receiving the transaction. To address the above issues, the relay node made some optimizations. First, Pioplat uses a TCP connection for transaction transmission instead of a time-consuming HTTP protocol to reduce users' submit transactions latency through remote procedure calls. Once any relay node receives a transaction from a user, it sends the complete transaction object directly to all its neighbors. It sends the complete transaction to other relays simultaneously, marking it as a user transaction. After receiving, all other relay nodes immediately send the transaction object to all their respective neighbors. Since transactions in permissionless blockchains are always public, whether in the transaction pool or on-chain, open to anyone. So regarding transmission, to save time, we do not use encryption to ensure confidentiality but only use SHA256-enabled HMAC with a secret salt to ensure the integrity of the message.

\begin{figure}[t]
\centerline{\includegraphics[width=0.5\textwidth]{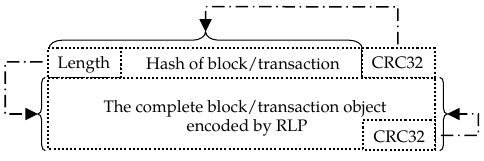}}
\caption{The datagram format in communication between relay nodes. It consists header and a complete block or transaction object.}
\label{fig_dataframe}
\end{figure}

\textbf{Communication in Pioplat:} From some point of view, relay nodes play a role as a gateway that helps users select neighbors from the blockchain p2p network. One of the challenges is making other nodes in the p2p network can not distinguish those relay nodes from normal full nodes, i.e., the behavior of relay nodes does not differ from normal full nodes in the view of other nodes. We achieve this by relying on an instrumented variant of the full node to get information. Such as fork id when the relay node handshakes with a new node and the complete block/transaction object when the relay node is requested by neighbors and can not find the corresponding data in its cache. Since the security communication between relay nodes and the full node is necessary, we use TLS 1.3, the latest version, to secure the communication. The duration of handshaking and redirecting requests have no impact on the latency of the relay node. However, for the communication between relay nodes, the transmission duration is part of the latency. We need to minimize both the bandwidth overhead and the computation overhead to reduce latency. TLS 1.3 is too costly to be used for communication between relay nodes. We implement a security and performance trade-off in the communication process between relay nodes. As shown in Fig.\ref{fig_dataframe}, the datagram consists of a header and a complete block/transaction object encoded by RLP. The length field in the header is 4 bytes long and stores the size of the complete block/transaction object encoded by RLP and used to cope with the sticky packet issue of TCP. The hash field in the header is 32 bytes long and is the hash of the block or transaction object in the packet. The two crc32 fields are 4 bytes long individually, and they act as weak HMAC to resist tampering with the header and the complete object respectively. We use the same key in all the relay nodes, each relay node generates the stream using AES-128-CTR cipher and caches the stream, then XOR each datagram with the stream. Since one compromised single relay node would compromise the entire system, we take the scheme above as an optional trade-off between security and performance. We will describe in subsection \ref{sub_optimization} how to avoid duplicated en-/decode and en-/decryption by the scheme above.

\textbf{Neighbor selection:} We use two hash tables to respectively record the timestamps of blocks and transactions received from each neighbor, no matter the complete objects or announcements. Every 10 minutes sort them in the order of precedence, then each node is scored two scores, namely block scores and transaction scores, with the node that relays quickly scoring higher. However, there is still a problem, if the relay node sends a transaction or block object/announcement to neighbors, those "late" neighbors who received the information would never send the transaction or block to the relay node. Thus those neighbors will be recognized as indifferent by neighbor selection and get negative scores, which may cause the relay node to disconnect from nodes other than those expected. Peri handles this issue by sampling 1/4 of all transactions for measurement\cite{tang2022strategic}. In Peri's method, nodes only send transactions with hashes divisible by four. We argue that this method can not address the issue well since 1/4 of transactions still have negative effects on scoring. In addition, only relaying 1/4 of transactions reduces the efficiency of information dissemination in the blockchain network. In our implementation, after the relay node receives a complete block/transaction object, the relay node waits for 1500ms first and then sends the object or announcement to its neighbors.

\subsection{Optimizations}
\label{sub_optimization}
\textbf{Remove duplicated en-/decode and en-/decryption:} As described above in communication datagram format, we attach block/transaction hash in 40 bytes header of each network packet. Whenever each relay node receives a message from other relay nodes, it first decrypts the first 40 bytes of the message using a stream key cached locally. The hash allows quick checking of whether the corresponding object exists in the local cache of the relay node. Thus replacing en-/decoding, en-/decryption, and duplicated computation hash of the object with memory copies, which saves duration.

\textbf{DDoS mitigation:} One of the key features of relay nodes is that they do not perform validity checks on received blocks/transactions. It does not matter if relaying invalid transactions to neighbors, but relaying invalid blocks will be actively disconnected. Therefore, Pioplat exposes a DDoS vulnerability, a malicious node can drastically reduce Pioplat's performance just by simply sending invalid blocks to relay nodes. We rely on Pioplat's full node to mitigate this issue. This can not prevent the attack in advance but significantly increases the cost to the attacker. Each relay node sends the block hash to the full node 3 seconds after receiving the block, and the full node responds whether it contains the block, if not, the relay node further sends the block to the full node to check the validity. If the block is invalid, the relay node disconnects from the neighbor that sent the invalid block and adds the neighbor to its blocklist.

\section{Evaluation}
Our evaluation answers the following questions.
\begin{itemize}
  \item Is the Pioplat system's latency further reduced when a relay node is added? (\ref{evaluation_subsection_B})
  \item Does Pioplat have an advantage over Peri in terms of latency reduction? (\ref{evaluation_subsection_C})
  \item How does the performance change when the parameter is tuned for different relay nodes? (\ref{evaluation_subsection_D})
  \item Can Pioplat make the sent transaction confirmed on-chain faster? i.e., let the transaction arrive at miners early. (\ref{evaluation_subsection_E})
\end{itemize}

\subsection{Experimental setup}
\label{evaluation_subsection_A}
We set up an Intel E-2286G, 128G RAM, 1.9T SSD bare metal cloud instance located in Tokyo, which was used to deploy the full node in the evaluation. We ran 29 experiments from November 20th to 30th. We used the Pioplat and a tailored bsc\cite{bnb2022bsc} during experiments. Those two programs logged timestamps and corresponding hashes to log files when receiving blocks or transactions. We interrupted programs at the end of each experiment, and log files were subsequently statistically analyzed using NumPy and Pandas\cite{harris2020array,reback2020pandas}. Some systemic biases should be taken into account. We refer to the bias reduction method adopted by Peri\cite{tang2022strategic}. First, adding each relay node's IP to its respective blocklist with an infinite expiration time to avoid the connection between them in the p2p network. In the meantime, we also included the full node's IP in the blocklist of all relay nodes to avoid the p2p connection. Thirdly, let the program make NTP requests periodically to alleviate the timestamp systematic error issue. We conduct the evaluation using the three methods below.

\begin{itemize}
    \item Baseline: We use version v1.1.17 of bsc, which was the latest version when we started the evaluation \cite{bnb2022bsc}. Set parameter \textit{maxPeers} to 200, which specifies the maximum number of neighbors to stay connected to, and the rest are default settings.
    \item Peri: The method proposed in \cite{tang2022strategic}, 50\% of neighbors are replaced every ten minutes. Like the Baseline, we also set the parameter \textit{maxPeers} to 200, which means that 100 nodes are kept in each round, and the rest are dropped.
    \item Pioplat: Method proposed in this paper, consisting of an instrumented variant of bsc as the full node, five relay nodes distributed in East Asia, North America, and Europe. We set the parameter \textit{maxPeers} of each relay node to 200.
\end{itemize}

To avoid the influence of geographical location, all three methods above mentioned collected data at the same place, Tokyo. To connect enough neighbors in the p2p network, each node in evaluation is allowed to warm up for three hours. After warming up, we run each node to record blocks/transactions received timestamps for further analysis. 

\begin{figure}[t]
\centerline{\includegraphics[trim={0.8cm, 0.5cm, 1.5cm, 1.9cm},clip,width=0.5\textwidth]{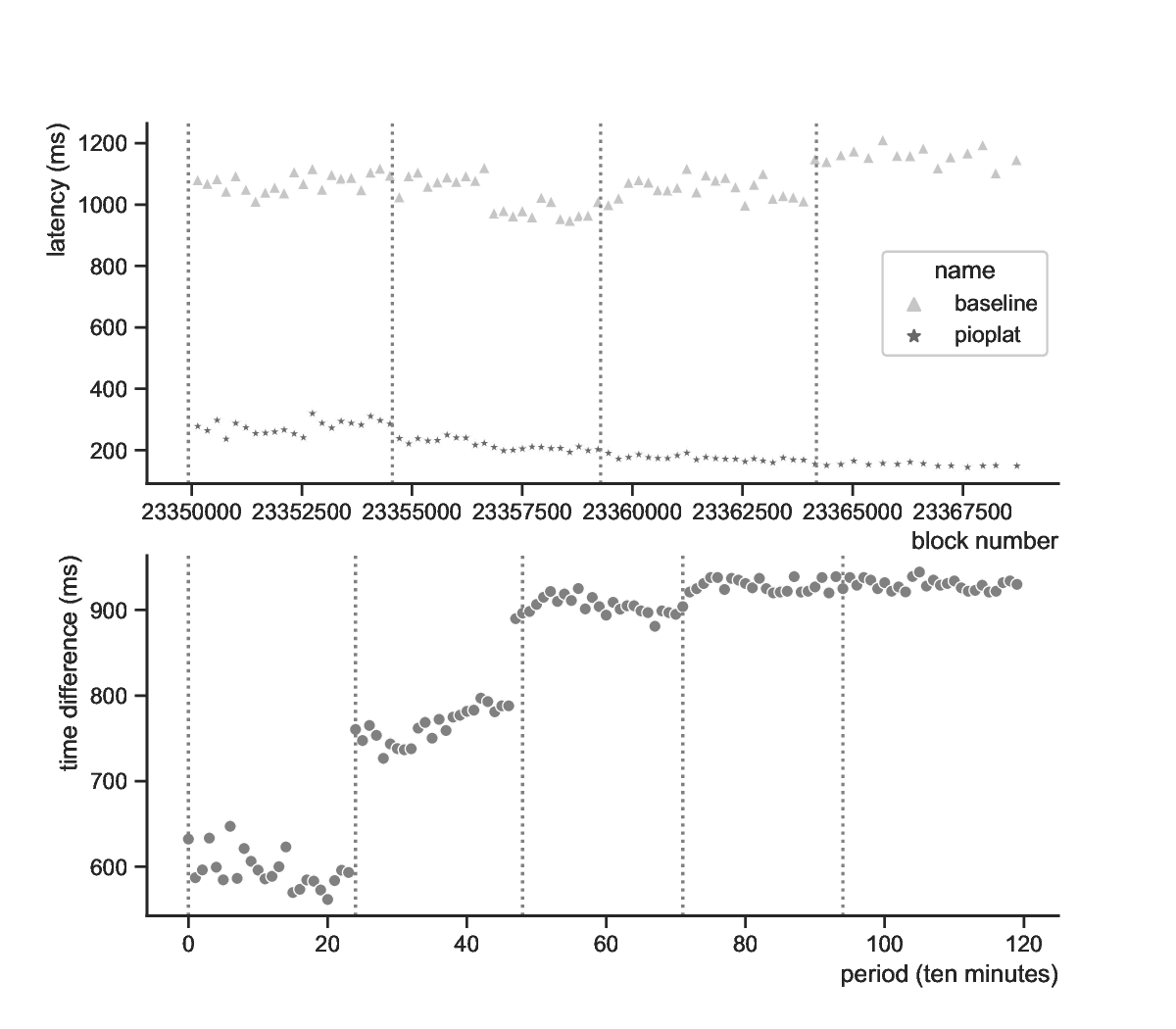}}
\caption{The top graph shows the latency of receiving blocks comparison between Baseline and Pioplat; The bottom graph shows the trend of Pioplat latency with the addition of a relay node, where each vertical dotted line indicates the moment when a new relay node is connected.}
\label{baseline_pioplat_fig}
\end{figure}

\subsection{Scalability of Pioplat}
\label{evaluation_subsection_B}
We run the Baseline node in Tokyo as a control. At the same time, we run five relay nodes in Tokyo, Singapore, New York, London, and Silicon Valley. When the warm-up phase is completed, these relay nodes are connected one by one sequentially every 4 hours. e.g., in the third four hours, three nodes, Tokyo, Singapore, and New York, are connected to each other. The data of the Baseline were collected on the full node, and the data of Pioplat were collected from another relay node server at the same location. Because the bsc uses PoS consensus, in which the block-producing interval is fixed, miners write the timestamp into a block when the block is generated. We subtract the timestamp when the block is received with the timestamp in the block to get the latency of receiving the block. The experimental result is shown in Fig.\ref{baseline_pioplat_fig}. For clarity, the latency shown in the Fig is the average latency for every two hundred blocks. The time difference shown in the Fig is to take the transactions received by both Baseline and Pioplat, and for each transaction, subtract the timestamp of Baseline from the timestamp of Pioplat, and take the average difference every ten minutes. It can be observed that the latency decreases significantly with connecting of the first three relay nodes and continues to decrease slightly with the addition of later relay nodes. After all five relay nodes are connected, Pioplat reduces the receiving block latency compared to the Baseline by roughly 800ms and reduces the latency of receiving transactions compared to the Baseline by roughly 920ms. Due to space limitations, we will not give a statistical graph of the percentage of transactions Pioplat first receives. We observed this percentage rise from 89\% to 98\% in the first 12 hours, then slowly increase to 99\%. This shows that the addition of the first three nodes significantly enhances performance. 

\begin{figure}[t]
\centerline{\includegraphics[trim={0.8cm, 0.5cm, 1.5cm, 1.9cm},clip,width=0.5\textwidth]{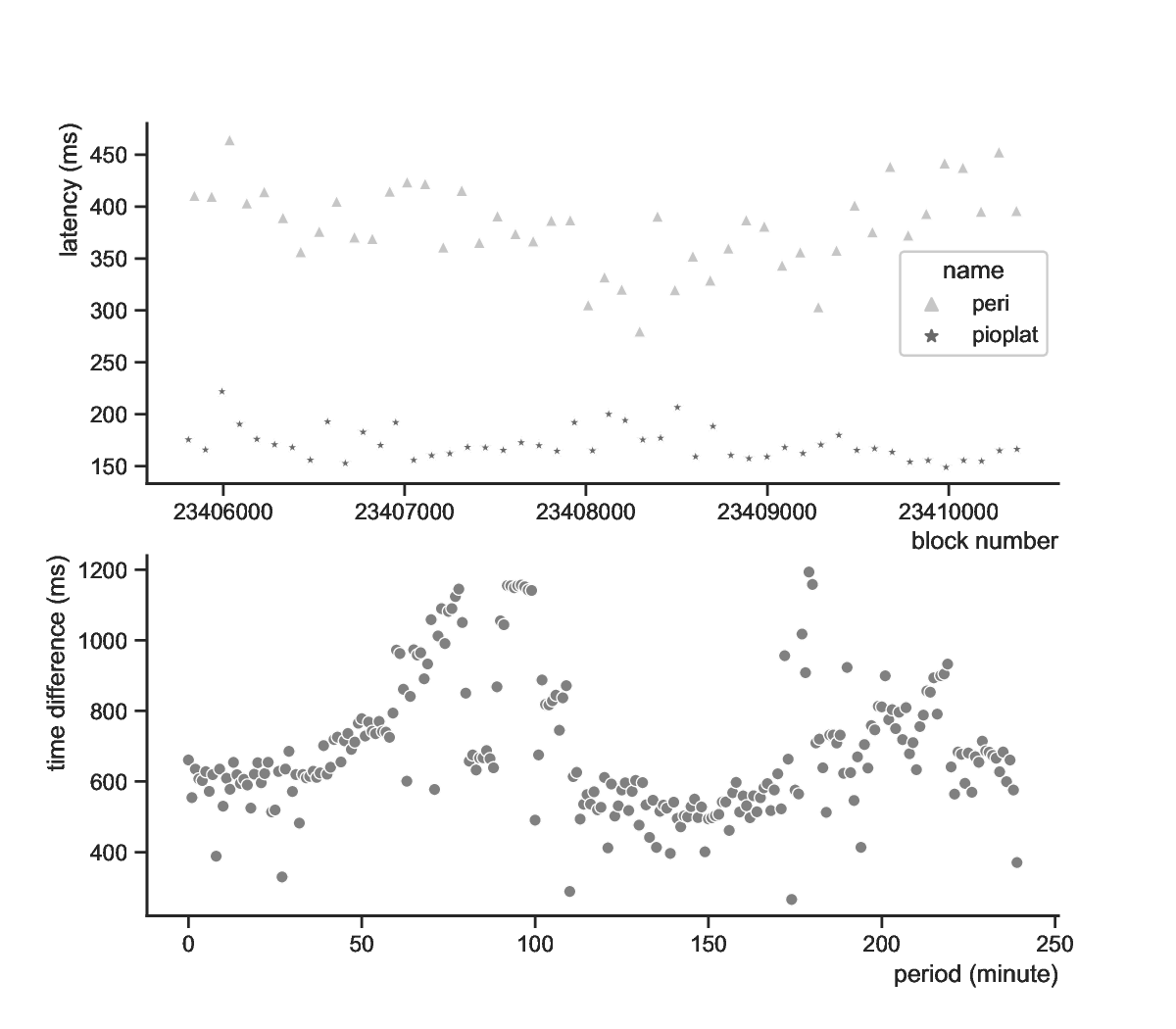}}
\caption{The top graph shows the latency of receiving blocks comparison between Peri and Pioplat; The bottom graph shows the timestamp difference of receiving transactions between Peri and Pioplat;}
\label{peri_pioplat_fig}
\end{figure}

\subsection{Compare with Peri}
\label{evaluation_subsection_C}
Like \ref{evaluation_subsection_B}, we run Peri as a control while directly keeping the five relay nodes connected as the Pioplat. Peri collected data from the full node, and Pioplat collected data from the relay node at the same location, Tokyo. After the warm-up phase, we start recording latency data for 4 hours. For clarity, we calculate the average latency every 100 blocks and the timestamp difference for receiving transactions every minute to demonstrate the performance. The experimental result is shown in Fig.\ref{peri_pioplat_fig}.
Regarding latency of receiving blocks, Peri is about 500ms lower than Baseline but still not as good as Pioplat, which runs five relay nodes and has a stable latency of 150ms to 200ms. We speculate that this is because the scoring rule in the neighbor selection strategy does not distinguish between the block announcement and the complete object, so Peri still has a high probability of receiving only announcements but not complete objects in a limited time and needs to request further. Regarding latency of receiving transactions, we found that 96\% of the transactions were received first by Pioplat. Although Pioplat's advantage over Peri is not as stable as the advantage over Baseline, it can still be about 600ms earlier in most moments.

\subsection{Tuning parameter}
\label{evaluation_subsection_D}
One of the advantages of Pioplat is customization, where users can tune the neighbor selection preference of relay nodes in different regions. During experiments in \ref{background_subsection_A} and \ref{background_subsection_B}, we set 30\% of slots reserved to neighbors that deliver blocks early and 50\% reserved to neighbors that deliver transactions early. We dumped the logs of the above experiments and noticed that more nodes deliver blocks early in North America and more nodes deliver transactions early in East Asia. This phenomenon is consistent with the conclusion of the prior work\cite{qiu2022geography}. Here we set the node in North America to keep a larger ratio of block neighbors and the nodes in East Asia to keep a larger ratio of transaction neighbors. We keep adjusting these parameters to find the optimal values. $ratio1 - ratio2$ in Tab.\ref{evaluation_table_1} represents that the relay node reserves $ratio1$ slots for neighbors delivering blocks quickly and $ratio2$ slots for neighbors delivering transactions quickly. Take $10\%-70\%$ as an example, which means that $ratio1=10\%$ and $ratio1=70\%$. We run the evaluation for three days, and each experiment first has a warm-up for 3 hours and then records timestamps for 4 hours. To reduce the temporal biases due to diurnal transaction traffic patterns, we repeated each experiment three times every 8 hours and took the average data. The results are shown in Tab.\ref{evaluation_table_1}. The \textit{block-arg} column indicates the latency of the relay node receiving blocks, and the \textit{tx-avg} column indicates the advantage of the relay node receiving transactions compared to the Baseline. The data in both columns are in milliseconds.

% \newcolumntype{H}{>{\setbox0=\hbox\bgroup}c<{\egroup}@{}}
\begin{table}[h]
    \caption{Neighbors selection preference parameters and results}
    \begin{center}
    \begin{tabular}{|c|c|c|c|c|}
    \hline
        \textbf{\textit{East Asia}} & \textbf{\textit{ID}} & \textbf{\textit{North America}} & \textbf{\textit{Block-avg (ms)}} & \textbf{\textit{Tx-diff (ms)}} \\ \hline
        \multirow{3}{*}{10\% - 70\%} & 1 & 70\% - 10\% & \textbf{182.64} & \textbf{1284.58} \\ \cline{2-5}
                                     & 2 & 60\% - 20\% & 183.48 & 1167.33 \\ \cline{2-5}
                                     & 3 & 50\% - 30\% & 197.30 & 1153.75 \\ \hline
        \multirow{3}{*}{20\% - 60\%} & 4 & 70\% - 10\% & \textbf{163.95} & \textbf{1207.73} \\ \cline{2-5}
                                     & 5 & 60\% - 20\% & 173.99 & 1186.74 \\ \cline{2-5}
                                     & 6 & 50\% - 30\% & 179.65 & 1187.15 \\ \hline
        \multirow{3}{*}{30\% - 50\%} & 7 & 70\% - 10\% & \textbf{153.48} & 983.93 \\ \cline{2-5}
                                     & 8 & 60\% - 20\% & 161.22 & 994.84 \\ \cline{2-5}
                                     & 9 & 50\% - 30\% & 170.56 & \textbf{1037.31} \\ \hline
    \end{tabular}
    \label{evaluation_table_1}
    \end{center}
\end{table}

The effect of tuning the parameter is not quite obvious, mainly on the magnitude of tens of milliseconds. Overall it is in line with our expectations. e.g., since we set more neighbors that deliver blocks quickly in relay nodes in North America, the first of the first three experiments has the lowest average block receipt delay. However, considering the round-trip-time from North America to Tokyo, experiment ID.7 performs better than ID.1. In terms of receiving transactions, unsurprisingly, thanks to Tokyo's relay node retaining more neighbors who deliver transactions quickly, ID.1 has the most significant advantage in receiving transactions.

\subsection{latency of sending transaction}
\label{evaluation_subsection_E}
Finally, we evaluate the advantage of Pioplat in sending transactions. Although one can send transactions in the test net for free, the number of miners on the test net is much less than in the mainnet, and the geographical distribution is quite different. Therefore, we evaluate directly on the mainnet. We run the Baseline located at Tokyo and the Pioplat setting as in experiment \ref{evaluation_subsection_A}. Given two transactions with different input data but the same nonce and the same gas price, the one that reaches the miner first will be packed into the block. So we submit the two transactions to the Baseline node and the relay nodes to check which transaction is finally confirmed on-chain. We repeated twice a minute for a total of 100 times. The result shows that the 94 confirmed transactions were sent via Pioplat, which powerfully demonstrates Pioplat's superior performance in reducing latency in sending transactions.

\section{Related Work}
\textbf{Comparing Pioplat and Peri/Perigee.} Pioplat presents a distributed system across different continents for latency reduction in the blockchain on top of the unstructured p2p network. It is tailored to low latency: multiple relay nodes receive data, and several optimizations reduce the overhead of delivering data between relay nodes. Peri/Perigee provided a p2p neighbor selection strategy based on timestamps of receiving transactions\cite{tang2022strategic,mao2020perigee}, which ignores reducing the latency of receiving blocks and transactions simultaneously. In addition, Peri adds the neighbors dropped in each round to a blocklist, which is not in line with the common churn of the p2p network. This will incorrectly block off many neighbors. Pioplat deploys an expirable blocklist to solve this issue.

\textbf{Structured p2p networks designed for low latency.} Existing blockchain is built on unstructured p2p networks, where nodes are randomly connected with each other\cite{kim2018measuring}. Swift is a solution that optimizes the P2P topology construction and broadcast algorithm in the structured network based on unsupervised learning and greedy algorithm\cite{wang2022data}. FRing leverages Intel SGX to build a geography-based P2P overlay network for fast and robust information broadcast in blockchain\cite{zhu2022design}. Urocissa provided a structured overlay protocol that maintains multiple minimum latency broadcasting trees, whereby each node communicates with its neighbors and makes decisions sovereignly to balance relaying tasks among participants\cite{zhu2022design}. All the above solutions are intrusive and are modifications of blockchain, and we emphasize that Pioplat is a non-intrusive framework that defers existing blockchain protocols.

\textbf{Targeted latency reduction for MEV.} Many profit opportunities can be observed in advance by participants. MEV refers to miners utilizing their ability to include, exclude, and reorder transactions arbitrarily for exploitation. This concept was first proposed by \cite{daian2020flash} and further be empirically studied in \cite{babel2021clockwork, zhou2021just, zhou2021high}. Babel studied how existing algorithms for latency optimization can be augmented with information about the transactions themselves to optimize peering algorithms \cite{babel2022strategic}. It also introduced an algorithm based on Peri, MEV-Peri, to maximize a node's expected MEV-weighted advantage from peer selection. The idea can be further applied to Pioplat to provide elastic, low-latency services for MEV.

\section{Conclusion}
We have presented a production-ready, feasible and customizable framework for latency reduction in the blockchain: Pioplat. It relies on relay nodes spanning different continents, equipped with a customizable p2p neighbor selection preference to reduce latency. Enabled by low-latency communication protocol between relay nodes, Pioplat allows users elastically add relay nodes, thus further reducing latency. We give the complete implementation and demonstrate its performance by comprehensive evaluation. Pioplat still has drawbacks, such as the ratio of neighbors who deliver blocks quickly and who deliver transactions quickly, which needs to be adjusted manually, and in the future, we will explore letting the relay node adjust this parameter automatically according to the blockchain clients' distruibution of the location.

% \section*{Acknowledgment}

% The preferred spelling of the word ``acknowledgment'' in America is without 
% an ``e'' after the ``g''. Avoid the stilted expression "one of us (R. B. 
% G.) thanks $\ldots$''. Instead, try "R. B. G. thanks$\ldots$". Put sponsor 
% acknowledgments in the unnumbered footnote on the first page.
\newpage
\bibliographystyle{IEEEtran}
\bibliography{IEEETran}

\begin{thebibliography}{10}
\providecommand{\url}[1]{#1}
\csname url@samestyle\endcsname
\providecommand{\newblock}{\relax}
\providecommand{\bibinfo}[2]{#2}
\providecommand{\BIBentrySTDinterwordspacing}{\spaceskip=0pt\relax}
\providecommand{\BIBentryALTinterwordstretchfactor}{4}
\providecommand{\BIBentryALTinterwordspacing}{\spaceskip=\fontdimen2\font plus
\BIBentryALTinterwordstretchfactor\fontdimen3\font minus \fontdimen4\font\relax}
\providecommand{\BIBforeignlanguage}[2]{{%
\expandafter\ifx\csname l@#1\endcsname\relax
\typeout{** WARNING: IEEEtran.bst: No hyphenation pattern has been}%
\typeout{** loaded for the language `#1'. Using the pattern for}%
\typeout{** the default language instead.}%
\else
\language=\csname l@#1\endcsname
\fi
#2}}
\providecommand{\BIBdecl}{\relax}
\BIBdecl

\bibitem{zhou2021high}
L.~Zhou, K.~Qin, C.~F. Torres, D.~V. Le, and A.~Gervais, ``High-frequency trading on decentralized on-chain exchanges,'' in \emph{2021 IEEE Symposium on Security and Privacy (SP)}.\hskip 1em plus 0.5em minus 0.4em\relax IEEE, 2021, pp. 428--445.

\bibitem{wang2022cyclic}
Y.~Wang, Y.~Chen, H.~Wu, L.~Zhou, S.~Deng, and R.~Wattenhofer, ``Cyclic arbitrage in decentralized exchanges,'' \emph{Available at SSRN 3834535}, 2022.

\bibitem{tang2022strategic}
W.~Tang, L.~Kiffer, G.~Fanti, and A.~Juels, ``Strategic latency reduction in blockchain peer-to-peer networks,'' \emph{arXiv preprint arXiv:2205.06837}, 2022.

\bibitem{babel2022strategic}
K.~Babel and L.~Baker, ``Strategic peer selection using transaction value and latency,'' in \emph{Proceedings of the 2022 ACM CCS Workshop on Decentralized Finance and Security}, 2022, pp. 9--14.

\bibitem{osipovich2020high}
A.~Osipovich, ``High-frequency traders push closer to light speed with cutting-edge cables,'' \emph{Wall Street Journal}, 2020.

\bibitem{osipovich2021high}
------, ``High-frequency traders eye satellites for ultimate speed boost,'' \emph{The Wall Street Journal. Apr}, vol.~1, 2021.

\bibitem{bloxroute_2022}
\BIBentryALTinterwordspacing
``Defi trading tools, mempool services, defi performance,'' Nov 2022. [Online]. Available: \url{https://bloxroute.com/}
\BIBentrySTDinterwordspacing

\bibitem{wang2021ethna}
T.~Wang, C.~Zhao, Q.~Yang, S.~Zhang, and S.~C. Liew, ``Ethna: Analyzing the underlying peer-to-peer network of ethereum blockchain,'' \emph{IEEE Transactions on Network Science and Engineering}, vol.~8, no.~3, pp. 2131--2146, 2021.

\bibitem{rohrer2019kadcast}
E.~Rohrer and F.~Tschorsch, ``Kadcast: A structured approach to broadcast in blockchain networks,'' in \emph{Proceedings of the 1st ACM Conference on Advances in Financial Technologies}, 2019, pp. 199--213.

\bibitem{mao2020perigee}
Y.~Mao, S.~Deb, S.~B. Venkatakrishnan, S.~Kannan, and K.~Srinivasan, ``Perigee: Efficient peer-to-peer network design for blockchains,'' in \emph{Proceedings of the 39th Symposium on Principles of Distributed Computing}, 2020, pp. 428--437.

\bibitem{silva2020impact}
P.~Silva, D.~Vavricka, J.~Barreto, and M.~Matos, ``Impact of geo-distribution and mining pools on blockchains: A study of ethereum,'' in \emph{2020 50th Annual IEEE/IFIP International Conference on Dependable Systems and Networks (DSN)}.\hskip 1em plus 0.5em minus 0.4em\relax IEEE, 2020, pp. 245--252.

\bibitem{cernera2022token}
F.~Cernera, M.~La~Morgia, A.~Mei, and F.~Sassi, ``Token spammers, rug pulls, and sniperbots: An analysis of the ecosystem of tokens in ethereum and the binance smart chain (bnb),'' \emph{arXiv preprint arXiv:2206.08202}, 2022.

\bibitem{noxx_2022}
\BIBentryALTinterwordspacing
Noxx, ``Dex arbitrage, mathematical optimisations,'' Aug 2022. [Online]. Available: \url{https://noxx.substack.com/p/dex-arbitrage-mathematical-optimisations}
\BIBentrySTDinterwordspacing

\bibitem{liu2022empirical}
Y.~Liu, Y.~Lu, K.~Nayak, F.~Zhang, L.~Zhang, and Y.~Zhao, ``Empirical analysis of eip-1559: Transaction fees, waiting time, and consensus security,'' \emph{arXiv preprint arXiv:2201.05574}, 2022.

\bibitem{mazorra2022price}
B.~Mazorra, M.~Reynolds, and V.~Daza, ``Price of mev: Towards a game theoretical approach to mev,'' in \emph{Proceedings of the 2022 ACM CCS Workshop on Decentralized Finance and Security}, 2022, pp. 15--22.

\bibitem{kim2018measuring}
S.~K. Kim, Z.~Ma, S.~Murali, J.~Mason, A.~Miller, and M.~Bailey, ``Measuring ethereum network peers,'' in \emph{Proceedings of the Internet Measurement Conference 2018}, 2018, pp. 91--104.

\bibitem{eth2014geth}
\BIBentryALTinterwordspacing
T.~E. Foundation, ``Go-ethereum,'' 2014. [Online]. Available: \url{https://github.com/ethereum/go-ethereum}
\BIBentrySTDinterwordspacing

\bibitem{bnb2022bsc}
\BIBentryALTinterwordspacing
T.~bsc~development team, ``Binance smart chain,'' 2020. [Online]. Available: \url{https://github.com/bnb-chain/bsc}
\BIBentrySTDinterwordspacing

\bibitem{ledger2020erigon}
\BIBentryALTinterwordspacing
Ledgerwatch, ``Erigon,'' 2020. [Online]. Available: \url{https://github.com/ledgerwatch/erigon}
\BIBentrySTDinterwordspacing

\bibitem{han2022using}
Y.~Han, D.~Xu, J.~Gao, and L.~Zhu, ``Using blockchains for censorship-resistant bootstrapping in anonymity networks,'' in \emph{International Conference on Information and Communications Security}.\hskip 1em plus 0.5em minus 0.4em\relax Springer, 2022, pp. 240--260.

\bibitem{matic2020bor}
\BIBentryALTinterwordspacing
maticnetwork, ``Bor,'' 2020. [Online]. Available: \url{https://github.com/maticnetwork/bor}
\BIBentrySTDinterwordspacing

\bibitem{wiki:User_Datagram_Protocol}
Wikipedia, ``{User Datagram Protocol} --- {W}ikipedia{,} the free encyclopedia,'' \url{http://en.wikipedia.org/w/index.php?title=User\%20Datagram\%20Protocol&oldid=1120834288}, 2022, [Online; accessed 03-December-2022].

\bibitem{mohammed2021hyperledger}
A.~H. Mohammed, A.~A. Abdulateef, and I.~A. Abdulateef, ``Hyperledger, ethereum and blockchain technology: A short overview,'' in \emph{2021 3rd International Congress on Human-Computer Interaction, Optimization and Robotic Applications (HORA)}.\hskip 1em plus 0.5em minus 0.4em\relax IEEE, 2021, pp. 1--6.

\bibitem{harris2020array}
\BIBentryALTinterwordspacing
C.~R. Harris, K.~J. Millman, S.~J. van~der Walt, R.~Gommers, and P.~Virtanen, ``Array programming with {NumPy},'' \emph{Nature}, vol. 585, no. 7825, pp. 357--362, Sep. 2020. [Online]. Available: \url{https://doi.org/10.1038/s41586-020-2649-2}
\BIBentrySTDinterwordspacing

\bibitem{reback2020pandas}
\BIBentryALTinterwordspacing
T.~pandas~development team, ``pandas-dev/pandas: Pandas,'' Feb. 2020. [Online]. Available: \url{https://doi.org/10.5281/zenodo.3509134}
\BIBentrySTDinterwordspacing

\bibitem{qiu2022geography}
H.~Qiu, T.~Ji, S.~Zhao, X.~Chen, J.~Qi, H.~Cui, and S.~Wang, ``A geography-based p2p overlay network for fast and robust blockchain systems,'' \emph{IEEE Transactions on Services Computing}, 2022.

\bibitem{wang2022data}
X.~Wang, X.~Jiang, Y.~Liu, J.~Wang, and Y.~Sun, ``Data propagation for low latency blockchain systems,'' \emph{IEEE Journal on Selected Areas in Communications}, 2022.

\bibitem{zhu2022design}
Y.~Zhu, C.~Hua, D.~Zhong, and W.~Xu, ``Design of low-latency overlay protocol for blockchain delivery networks,'' in \emph{2022 IEEE Wireless Communications and Networking Conference (WCNC)}.\hskip 1em plus 0.5em minus 0.4em\relax IEEE, 2022, pp. 1182--1187.

\bibitem{daian2020flash}
P.~Daian, S.~Goldfeder, T.~Kell, Y.~Li, X.~Zhao, I.~Bentov, L.~Breidenbach, and A.~Juels, ``Flash boys 2.0: Frontrunning in decentralized exchanges, miner extractable value, and consensus instability,'' in \emph{2020 IEEE Symposium on Security and Privacy (SP)}.\hskip 1em plus 0.5em minus 0.4em\relax IEEE, 2020, pp. 910--927.

\bibitem{babel2021clockwork}
K.~Babel, P.~Daian, M.~Kelkar, and A.~Juels, ``Clockwork finance: Automated analysis of economic security in smart contracts,'' \emph{arXiv preprint arXiv:2109.04347}, 2021.

\bibitem{zhou2021just}
L.~Zhou, K.~Qin, A.~Cully, B.~Livshits, and A.~Gervais, ``On the just-in-time discovery of profit-generating transactions in defi protocols,'' in \emph{2021 IEEE Symposium on Security and Privacy (SP)}.\hskip 1em plus 0.5em minus 0.4em\relax IEEE, 2021, pp. 919--936.

\end{thebibliography}

\end{document}